\documentclass[%
 reprint,
 amsmath,amssymb,
 aps,
]{revtex4-2}

\usepackage{graphicx}
\usepackage{dcolumn}
\usepackage{bm}

\usepackage{parskip}
\usepackage{hyperref}

\begin{document}

\preprint{APS/123-QED}

\title{Effect of Spin Fluctuations on Magnetoresistance and Anomalous Hall Effect in the Chiral Magnet Co$_{8}$Zn$_{8}$Mn$_{4}$}

\author{P. Saha}
\affiliation{School of Physical Sciences, Jawaharlal Nehru University, New Delhi-110067, India}
\author{P. Das}
\affiliation{School of Physical Sciences, Jawaharlal Nehru University, New Delhi-110067, India}
\author{M. Singh}
\affiliation{School of Physical Sciences, Jawaharlal Nehru University, New Delhi-110067, India}
\author{R. Rai}
\affiliation{School of Physical Sciences, Jawaharlal Nehru University, New Delhi-110067, India}

\author{S. Patnaik}
\email{spatnaik@jnu.ac.in}
\affiliation{School of Physical Sciences, Jawaharlal Nehru University, New Delhi-110067, India}

\date{\today}

\begin{abstract}
The $\beta$ -Mn type cubic Co-Zn-Mn alloys have received significant attention recently due to their ability to host skyrmions at room temperature. The objective of this study is to thoroughly investigate the magnetotransport properties of one member of the family (Co$_{8}$Zn$_{8}$Mn$_{4}$) and delineate the role played by different scattering mechanism in the presence of topological invariance and spin fluctuations. A negative magnetoresistance is observed over a  wide temperature range of 50K to 300K. The deviation of isothermal magnetoresistance curves from linearity (for T$<<$T$_\text{c}$) to non-linearity (for T$<$T$_\text{c}$ and T$\sim$T$_\text{c}$), points towards the transition from the dominance of magnons to spin fluctuations. In the paramagnetic phase, the change in the shape of the magnetoresistance curve has been explained using the Khosla and Fischer model. The relationship between the anomalous Hall effect and longitudnal resistivity reveals the dominance of the skew-scattering mechanism, which is inexplicable based on the theories of semiclassical magnetotransport. We experimentally determine that the spin fluctuation is the source of the skew-scattering mechanism in Co$_{8}$Zn$_{8}$Mn$_{4}$. Skew-scattering mechanisms generally predominate in compounds with high conductivity. But our findings demonstrate that there is sufficient ground for exceptions in skyrmionic alloys. Our work throws new light on the dominance of varied scattering mechanisms in chiral magnets with low conductivity.
\begin{description}
\item[PACS numbers]
75.70.Ak, 72.10.Di
\item[Keywords]
Magnons, Spin fluctuations, Magnetoresistance, Anomalous Hall effect, Skyrmions
\end{description}
\end{abstract}

\maketitle


\section{\label{sec:level1}INTRODUCTION}

The skyrmionic phases of chiral magnets are of significant current interest, and relevance, at present \cite{r37}. From the theoretical perspective, it is important to understand the mechanism of magnetotransport in such topologically protected spin textures. More so for technological reasons, where their projected usage as magnetic storage devices, based on 5 orders of magnitude smaller drive current requirement compared to domain wall movement of ferromagnets, is of revolutionary potential. Of particular interest are the non-centrosymmetric ferromagnetic alloys with extremely rich magnetic phase diagrams that are inclusive of stable skyrmionic phases \cite{r1,r2,r3,r4}. In such systems, the competition between the Dzyaloshinskii Moriya (DM) interaction, associated with the inversion broken symmetry, and the ferromagnetic exchange interaction, tends to stabilize the skyrmionic phase just below the transition temperature (T$_{\text{c}}$). Recently, a new group of $\beta$ -Mn type chiral Co-Zn-Mn alloys have come to the fore that host the skyrmions near or above room temperature. This provides a considerable leap towards practical applications \cite{r7,r8}.  \newline The Co-Zn-Mn ((Co$_{0.5}$Zn$_{0.5}$)$_{20-x}$Mn$_{\text{x}}$,  with 0$<$x$<$6) family of compounds hosts intricate magnetic phase diagrams and its transition temperature (T$_{\text{c}}$)  as well as its magnetic properties can be tuned by the concentration of Mn(x) \cite{r7,r8}. The first member of this family (x=0), Co$_{10}$Zn$_{10}$, exhibits a helimagnetic phase below T$_{\text{c}}\sim$  415K \cite{r7}. This T$_{\text{c}}$ drops dramatically with increasing Mn substitution and ultimately reaches $\sim$~160K  for Co$_{7}$Zn$_{7}$Mn$_{6}$ \cite{r9}. More intriguingly, among the family of Co-Zn-Mn compounds, Co$_{8}$Zn$_{8}$Mn$_{4}$  hosts metastable skyrmions that exist in a wide temperature and field range \cite{r10}  unlike the narrow temperature range that exists below T$_{\text{c}}$  in other members of this family. This further broadens its potential application in skyrmionic memory devices \cite{r6}.  
\newline The electrical transport behavior of the skyrmionic systems holds a pivotal role towards its potential application in spintronics \cite{r38}. In this work, we focus on identifying the scattering mechanisms involved in shaping the magnetotransport behaviour as well as the role played by them in influencing the temperature dependence of anomalous Hall effect (AHE) and magnetoresistance (MR) in single crystals of Co$_{8}$Zn$_{8}$Mn$_{4}$ .
AHE is of prime importance towards utilization of ferromagnets in varied technical applications \cite{r11}. Despite intensive research, the anomalous Hall effect (AHE) has persistently remained a controversial matter in Co-Zn-Mn alloys \cite{r12,r13}.  While it is well understood that the ordinary Hall effect (OHE) stems from the deflection of charge carriers due to the externally applied magnetic field, the phenomological origin of AHE still poses open questions \cite{r11,r12,r13,r14,r15}. The AHE’s  distinctive variation with temperature for different magnetic materials also poses challenge for its correct interpretation \cite{r16,r17,r18}. In general, three different mechanisms participate in shaping the behaviour of AHE, i.e., the extrinsic skew scattering \cite{r19,r20}, the side jump mechanism \cite{r21}, and the intrinsic Karplus Luttinger mechanism \cite{r22}. With regard to Co-Zn-Mn compounds, some reports indicate the dominance of extrinsic skew scattering mechanism \cite{r12} while the other reports imply the dominance of intrinsic mechanism such as in Co$_{9}$Zn$_{9}$Mn$_{2}$ \cite{r13}. Furthermore, the longitudinal conductivity ($\sigma$$_{\text{xx}}$) of ferromagnets has also been broadly classified  into three regimes on the basis of the dominance of different mechanisms on AHE \cite{r11}. According to this classification, skew scattering contribution dominates the AHE when the material exhibits a very high longitudinal conductivity i.e. $\sigma_{\text{xx}}$ $>$ 10$^{6}$ ($\Omega$ cm)$^{-1}$ \cite{r11}. The conductivity of Co-Zn-Mn compounds falls in the range $3\times10^{3}$$(\Omega cm)^{-1}$ $<$ $\sigma_{\text{xx}}$ $<$ $9\times10^{3}$($\Omega$ cm)$^{-1}$ which is considerably lower than the conductivity required for the dominance of the skew scattering mechanism. This has put the results associated with anomalous Hall resistivity of Co$_{7}$Zn$_{8}$Mn$_{5}$  \cite{r12} under question. Moreover, a theoretical report by Ishizuka et al. has proposed that the Skew scattering mechanism can stem from the fluctuating but locally correlated spins \cite{r23}. This kind of unconventional behaviour has been experimentally verified in SrCoO$_{3}$ thin films, where the skew scattering contribution originated from the spin fluctuations with impurity induced local inversion symmetry breaking \cite{r24}.  It is known that Co$_{8}$Zn$_{8}$Mn$_{4}$ crystallizes in cubic $\beta$ -Mn type structure, where broken inversion symmetry results in antisymmetric DM interaction. Therefore, it would be interesting to investigate how spin fluctuations affect magnetotransport characteristics of low conductivity chiral magnets such as Co$_{8}$Zn$_{8}$Mn$_{4}$. \newline In this paper, we discuss the experimental results on the scattering mechanisms which tend to influence the magnetotransport behaviour in  single crystal Co$_{8}$Zn$_{8}$Mn$_{4}$. We study the dominance of different scattering mechanisms in different temperature regimes and ascertain that the spin fluctuations play a predominant role in shaping the temperature dependence of AHE. Our objective is to draw definitive conclusions on the microscopic origin of magneto-electronic transport in chiral ferromagnetic phases of Co-Zn-Mn alloys\cite{r12,r13}.

\section{EXPERIMENTAL TECHNIQUES}
Single crystals of Co$_{8}$Zn$_{8}$Mn$_{4}$ were prepared using the modified Bridgman technique. Stoichiometric amounts of Cobalt , Zinc  and Manganese were ground for half an hour using a mortar and pestle. This mixture was then vacuum sealed in a quartz tube and heated to 1000$^{\circ}$C for 15 hours. After that, it was slowly cooled down to 700$^{\circ}$C over a period of 8 days followed by a water quench. An ingot with large single crystalline grains (Inset (ii) fig. 1(a))   was obtained as an end product. The phase purity and crystal structure were identified using Rigaku x-ray diffractometer (Miniflex-600). Bruker x-ray diffractometer was used to corroborate the single-crystalline nature of our sample. The low temperature magnetotransport measurements were conducted using a Cryogenic Cryogen Free Magnet (CFM) in conjunction with variable temperature insert (1.6 K, 8 Tesla). Energy dispersive x-ray spectroscopy (EDAX) analysis was achieved using a Bruker AXS microanalyzer.
\begin{figure}[htp]
\includegraphics{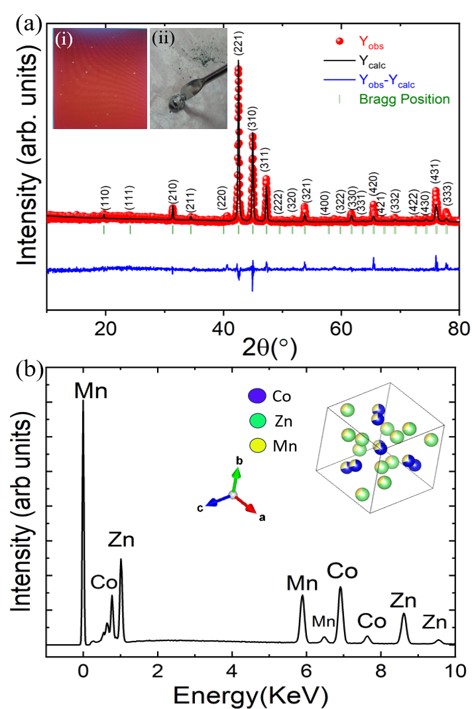}
\caption{\label{figure1}(a) X-Ray diffraction pattern of powdered single crystal of Co$_{8}$Zn$_{8}$Mn$_{4}$. Inset (i) shows the spots in the Laue diffraction pattern. Inset (ii) shows the image of the single crystalline ingot of Co$_{8}$Zn$_{8}$Mn$_{4}$. (b) depicts the EDX pattern for Co$_{8}$Zn$_{8}$Mn$_{4}$. Inset of (b) shows the crystal structure of $\beta$ -Mn type Co-Zn-Mn alloys}
\vspace{-5mm}
\end{figure}

\section{RESULTS AND DISCUSSIONS}

\subsection{\label{sec:level2}Structural Characterisation}
Fig 1(a) shows the Rietveld refined X-Ray diffraction pattern of crushed single crystals of Co$_{8}$Zn$_{8}$Mn$_{4}$. All the observed peaks match well with the spacegroup of P4$_{1}$32 (no. 213). The lattice parameters derived from the  refinement are a = b = c = 6.377 \AA{} which is in accordance with the cubic phase of Co-Zn-Mn alloys and in agreement with the previous reports \cite{r25} . Moreover, the spots in the Laue diffraction pattern (fig1a. inset(i)) verifies the single crystal nature of our sample. Inset of fig 1(b) shows  the  schematic diagram of a unit cell of Co$_{x}$Zn$_{y}$Mn$_{z}$  (x+y+z= 20) compound which reveals a cubic crystal structure. The unit cell consists in total of 20 atoms which are located in two types of crystallographic sites: 8c and 12d. The 8c sites which exhibit a distorted diamond lattice are more likely to be occupied by the Co atoms \cite{r28}. The "hyperkagome" lattice is created by the connections between the neighbouring 12d sites, which is made up of three dimensional corner sharing triangles \cite{r28}. These 12d sites are mostly preferred by the Mn and Zn atoms \cite{r25} . The EDAX analysis was done on various points of the sample surface and the elemental composition was found to be close to the intended stoichiometric ratio.
\begin{figure}[!t]
\includegraphics{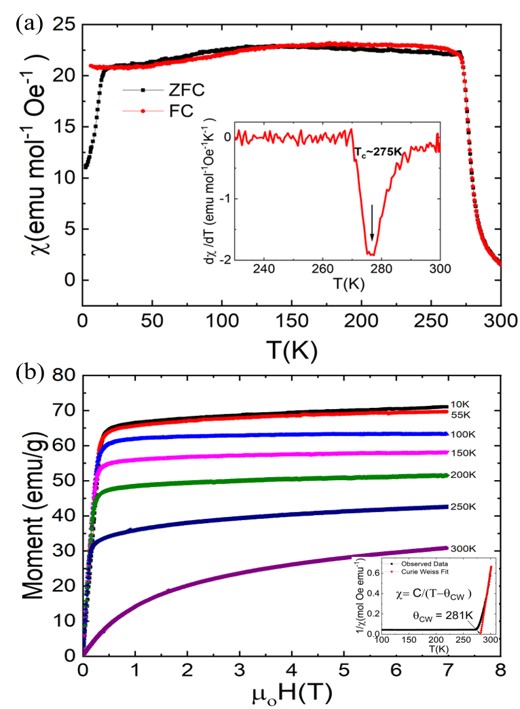}
\caption{\label{figure2} (a) shows the temperature dependent dc susceptibility in zero field cooled warming (ZFC) and field cooled warming (FC) protocols where the field is applied perpendicular to the sample plane. Inset (i) shows temperature dependence of $d\chi/dT$ vs $ T $  with a minima at $T_{c}$ = 275K . (b) shows the isothermal field dependent magnetization behaviour at different temperatures for field applied perpendicular to the sample plane. Inset (i) black curve shows the change in inverse susceptibility 1/$\chi$ with temperature. Also included is an extrapolation using the Curie Weiss equation.}
\end{figure}

\subsection{Magnetization}%
The temperature dependent change in dc susceptibility in the zero field cool warming (ZFC) and field cool warming (FC) protocols with an applied field of 100 Oe is shown in the main panel of fig.2a. The ZFC and FC curves depict an abrupt drop in the behaviour above T$\sim$270K which is a clear indication of a ferro-para magnetic phase transition. The Curie temperature (T$_{c}$) is obtained through the pronounced drop in the temperature derivative of dc susceptibility, which is at 275K (inset of fig.2a). Similar to earlier publications \cite{r12,r13,r25,r26}, a slight bifurcation between ZFC and FC has been detected just below T$_{c}$. This bifurcation results from different magnetization caused by aligned and misaligned domains in the FC and ZFC cases. Further, a smooth decrease in the susceptibility can be seen below 120K, which is a common feature in the Mn rich Co-Zn-Mn alloys \cite{r7}. This is due to the disorder created by the antiferromagnetically correlated Mn spins. Below 20K, a sudden drop is observed in the ZFC curve. This is because at low temperatures, the Mn moments freeze into a fully disordered re-entrant spin glass state \cite{r27}. This kind of dip has been observed in other Co$_{x}$Zn$_{y}$Mn$_{z}$ compounds as well \cite{r28}. Fig 2b inset (ii) shows the variation of the inverse dc susceptibility with temperature. The linear region has been fitted using the Curie Weiss equation:
\begin{equation}
    \chi=\frac{C}{T-\Theta_{cw}}
\end{equation}
                 
Here $\Theta_{cw}$  is the Curie Weiss temperature and C is the Curie constant from which the effective moment ($\mu_{\text{eff}}$) can be obtained. The positive value of $\Theta_{cw}$ at 281K reconfirms the ferromagnetic nature of Co$_{8}$Zn$_{8}$Mn$_{4}$. Fig 2(b) shows the field dependent magnetization behaviour at various temperatures. Here, the saturated magnetization rises continuously with decreasing temperature and finally reaches a maximum value of 71emu g$^{-1}$ at 10K. The steep increase in magnetization followed by saturation is reflective of the ferromagnetic order.

\subsection{Resistivity}%
Fig 3(a) shows the temperature dependent change in longitudinal resistivity ($\rho_{\text{xx}}$) at zero field. The increase in resistivity with temperature clearly indicates the metallic character of this sample. At low temperatures, a resistivity minima is observed around 20K. The obtained data indicate initial decrease in resistivity followed by a gradual increase after 20K with increasing temperature. This nature of resistivity with evolving temperature evidences the presence of a fully disordered re-entrant spin-glass phase. In Co$_{x}$Zn$_{y}$Mn$_{z}$ compounds, ferromagnetic Co moments along with the large dynamically disordered Mn moments exhibit a two-sublattice magnetic behaviour \cite{r25}. At low temperatures, the large dynamically disordered Mn moments freeze into a spin glass state while the Co spins continue to be ferromagnetically ordered \cite{r25}. This behaviour is consistent with the low temperature region of the ZFC curve (fig.2a). Similar behaviour has also been observed in Co$_{7}$Zn$_{8}$Mn$_{5}$ \cite{r12}. This spin glass state is related to the concentration of the Mn atoms, and this phase is absent in Mn-deficient compositions of Co-Zn-Mn alloys \cite{r13}. Above 275K, there is a gradual change in slope, which points towards the ferromagnetic phase transition. 
As stated by Matthiessen’s rule, different scattering mechanisms contribute to the longitudinal resistivity ($\rho_{\text{xx}}$)  of a metallic sample. This is expressed by the following equation:
\begin{equation}
    \rho_{\text{total}}=\rho_{\text{res}}+\rho_{\text{e-e}}(T)+\rho_{\text{e-m}}(T)+\rho_{\text{e-ph}}(T)
\end{equation}                                                                     Here, the first term ($\rho_{\text{res}}$) is the residual resistivity, which is a temperature-independent indicator of lattice imperfections and impurities. It is worth noting that the residual resistivity of our sample is 111$\mu\Omega$-cm which is much less than the previous reported results \cite{r12,r13} . The second ($\rho_{\text{e-e}}$), third ($\rho_{\text{e-m}}$) and fourth ($\rho_{\text{e-ph}}$) terms are the contributions that arise from electron-electron, electron-magnon and electron-phonon scattering respectively. The $\rho_{\text{e-e}}$ and $\rho_{\text{e-m}}$ terms have a quadratic dependence on temperature whereas $\rho_{\text{e-ph}}$ has a linear dependence on temperature. As shown in the fig.3(a), the dominant scattering mechanisms have been extracted for different temperature ranges in the ferromagnetic phase. The resistivity between the temperature range of 30K-110K clearly fits the quadratic dependence to temperature (T$^2$). This reflects the dominance of electron-electron or electron-magnon scattering in this regime. It is to be noted that electron-magnon scattering is a strongly field-dependent phenomenon and its dominance on the longitudinal resistivity can be assured through the analysis of the field- dependent longitudinal resistivity measurements, as will be seen in the later section. In the higher temperature region (112K-265K), the curve shows a sum of linear and quadratic dependence on temperature. It is evident from this fit that at higher temperatures in the ferromagnetic phase, electron phonon scattering also contributes along with electron magnon scattering. 

\begin{figure}[!t]
\includegraphics{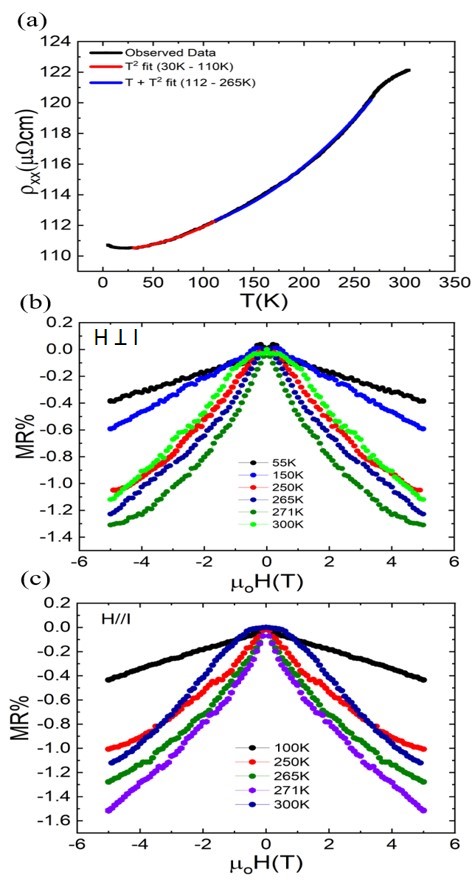}
\caption{\label{fig:epsart} (a) shows the temperature dependent resistivity from 5K to 304K and the theoretical fitting in different ranges of temperature. (b) and (c) shows the isothermal magnetoresistance curves for fields (upto 5T) applied perpendicular and parallel to the sample plane.}
\end{figure}
\subsection{Magnetoresistance}%
In a typical ferromagnet, the field driven magnetoresistance (MR) is substantially affected by the suppression of magnons and spin fluctuations. Magnons are low energy and long wavelength spin wave modes which dominate the transport properties of the magnetic materials. At zero temperature, the energy gap ($\Delta$)  between the majority and minority spin bands is large and this tends to prevent the spin flip transition. However, as the magnon wavevector (q) increases, the band gap $\Delta$ decreases and this leads to enhancement in electron-magnon scattering. At intermediate temperatures, the magnon wave vector q reaches a certain value (q$_{\text{c}}$) where the magnon dispersion curve enters the Stoner excitation continuum \cite{r29}. As a consequence, these low frequency propagating spin waves (magnons) transition to non-propagating spin fluctuations. Hence, at very low temperatures (T$<<$T$_{c}$), the resistivity behaviour is dominated by the electron-magnon scattering and at intermediate (T$<$T$_{c}$) and high temperatures (T$\sim$T$_{c}$), the spin fluctuations may play a dominant role on the transport properties \cite{r29}. 
The isothermal magnetoresistance measurements were performed with field applied parallel and perpendicular to the sample plane. The Hall resistivity contribution was removed from the transverse and longitudinal MR using the following equation:
\begin{equation}
    \rho_{\text{xx}}(\mu_{o} H,T)=\frac{\rho_{\text{xx}}(\mu_{0} H,T)+\rho_{\text{xx}}(-\mu_{0} H,T)}{2}
\end{equation}
\begin{figure}[b]
\includegraphics{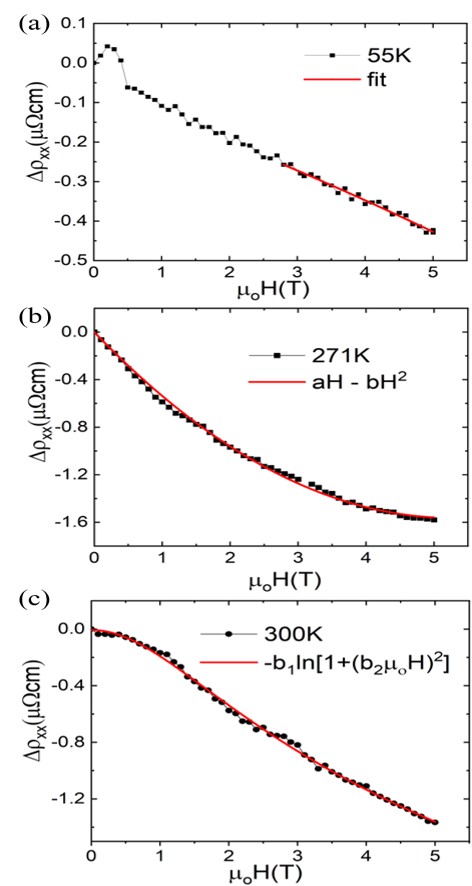}
\caption{\label{fig:epsart} (a) black curve shows the change in resistivity with field at 55K whereas the red curve shows the fit corresponding to equation (5). (b) black curve shows the field dependent change in resistivity at 271K and the red curve shows the fitting using equation (6). (c) black curve shows the observed variation in resistivity as a function of field in the paramagnetic regime (300K) and the red curve shows the theoretical fit to this curve using equation (7).}
\end{figure}
       
Fig.3(b) and fig.3(c) depicts the transverse and longitudinal MR with field upto 5T. The MR\% was obtained using the equation shown below:
\begin{equation}
    MR\%=\frac{\Delta\rho_{xx}}{\rho_{xx}(T,0)}\times100=\frac{\rho_{\text{xx}}(T,\mu_{0} H)-\rho_{\text{xx}}(T,0)}{\rho_{xx}(T,0)}\times100
\end{equation}
where $\rho_{\text{xx}}(T,\mu_{0} H)$ and $\rho_{\text{xx}}(T,0)$ are the resistivities at non-zero and zero fields. Fig.3(b) and fig.3(c) illustrate the increasing negative MR\% with evolving temperature up until T approaches T$_{c}$. The shape of the curve also tends to change for different temperature ranges. 
As seen in fig.3(b) and fig.3(c), for low temperature there is a sharp rise in the MR at low fields followed by a linearly negative behaviour. This sharp increase in MR is due to the presence of scattering from domain walls. This domain wall scattering is very evident from the fact that the value of field upto which this behaviour is observed in MR coincides with the field value above which the magnetization saturates (fig.2(b) shows field dependent magnetization). 
In the low temperature region (fig.3(b) and fig.3(c)), the MR shows a negative and nonsaturating linear behaviour. The qualitative explanation of the observed negative magnetoresistance is the following. The application of magnetic field, enhances the gap in the magnon spectrum, and this results in a decline in the electron magnon scattering leading to a negative MR\%. As already discussed, the magnon population tends to increase which increasing temperature and decrease with increasing field. So the decline in magnon population due to an applied field is overcompensated at high temperatures resulting in higher negative MR\%. Raquet et al \cite{r30}  derived an expression to describe the negative change in resistivity with applied field due to the suppression in spin flip electron-magnon scattering. This equation is valid for fields below 100T and for a temperature range of T$_{c}$/5 to T$_{c}$/2:   
\begin{equation}
    \rho_{\text{xx}}(T,B)\propto \frac{BT}{D(T)^2}\ln(\frac{\mu_{B} B}{k_{B} T})
\end{equation}
where D(T) is defined as magnon stiffness or magnon mass renormalization, B is magnetic field and T is the temperature. Fig.4(a) shows that the above equation fits well for data between 3 to 5T (at 55K). This verifies that the field driven decrease in resistivity is due to the suppression in magnon population.
As seen in fig.3(b) and fig.3(c), for intermediate (T$<$T$_{\text{c}}$)  and high temperatures (T$\sim$T$_{\text{c}}$), the shape of the MR curve deviates from linearity, although the trend of increasing negative MR with increasing temperature is still maintained. This kind of MR indicates towards the dominance of exchange-enhanced spin fluctuations \cite{r30,r31} in high temperature regime. With evolving field the magnetic excitations due to spin fluctuations get suppressed and this results in a negative MR. This decrement in resistivity with increasing field due to the dampening of spin fluctuations can be described by the following equation \cite{r31} :
\begin{equation}
    \Delta\rho_{\text{xx}}=bH^2-aH
\end{equation}                                              The above equation is obeyed well by the magnetoresistance curves of intermediate and high temperature regime (T$\ge$200K) as seen in fig.4(b). This confirms the dominance of spin fluctuations in this temperature range. It is to be noted that for T $\le$150K, this equation did not fit well. This is due to the dominance of magnons over spin fluctuations in this regime.
At 300K (T$>$T$_{\text{c}}$), in the paramagentic phase, the MR curve exhibits a positive curvature like behaviour at low fields which progresses into a negative MR. This behaviour could be a consequence of the persistence of local moments that contribute to paramagnetic susceptibility \cite{r16}. Khosla and Fischer have reported this kind MR in the In doped CdS samples which arise due to scattering from local moments \cite{r32}. They have subsequently developed the following semi-empirical relation which comprises of a degenerate electron gas system with local moments that undergoes thermal fluctuations \cite{r32} 
\begin{equation}
    \Delta\rho_{\text{xx}}=-b_{1}\ln[1+(b_{2}\mu_{0}H)^2]
\end{equation} 
      
 This equation have been applied to many material systems \cite{r16,r18} . Fig.4(c) shows that the MR at 300K can be fitted using equation(7). The parameters derived from this fit are b$_{1}$= 0.58$\mu$$\Omega$-cm and b$_{2}$= 0.62m$^{2}$/Vs. It is worth noting that this equation (7) could not fit the MR behaviour at T$<$T$_{c}$ and this clearly indicates the phase transition between these two temperature regimes as well as validates the involvement of different scattering processes in the two temperature ranges.
 \begin{figure}[b]
\includegraphics{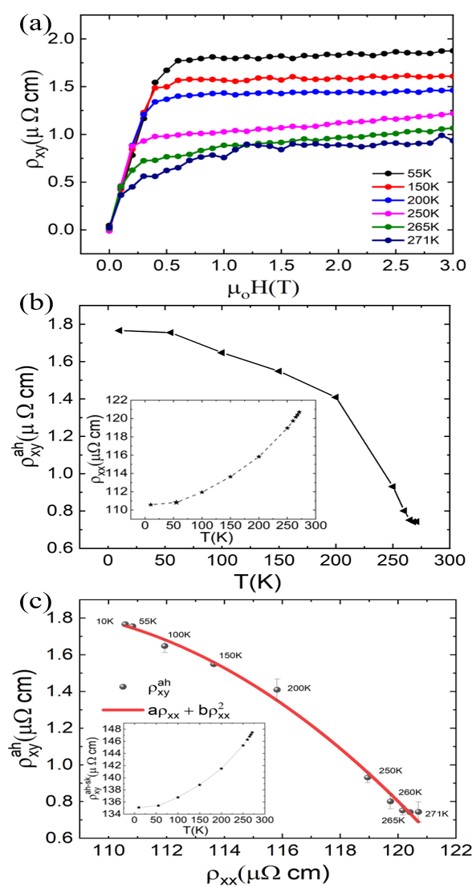}
\caption{\label{fig:epsart}(a) shows the isothermal Hall resistivity curves for fields upto 5T. (b) shows the temperature dependence of anomalous Hall resistivity ($\rho_{\text{xy}}^{\text{ah}}$) with temperature. Inset of (b) shows the variation of $\rho_{xx}$ with temperature. (c) shows the change in anomalous Hall resistivity ($\rho_{\text{xy}}^{\text{ah}}$) as a function of longitudinal resistivity ($\rho_{xx}$) and the red curve shows the scaling fit using equation. Inset of (c) shows the temperature dependent variation of skew scattering contribution ($\rho_{\text{xy}}^{\text{ah-sk}}$).}
\end{figure}

\subsection{Anomalous Hall Effect}%
Anomalous Hall effect (AHE) has emerged as the smoking gun for the revelation of new physics and detection of nontrivial spin textures. In general, the origin of AHE can be explained by the contributions of three different mechanisms. One of them is the extrinsic skew scattering mechanism, which results from the asymmetric scattering of up and down spins in the presence of spin orbit interaction \cite{r19,r20}. The skew scattering contribution is directly proportional to longitudinal resistivity ($\rho_{\text{xx}}$). The second is the side jump mechanism, which is of extrinsic origin and arises from the crosswise transference of the wave packet’s centre of mass during the process of scattering \cite{r21} . The third is the intrinsic Karplus Luttinger (KL) mechanism which is a scattering free contribution and is mainly associated with the Berry curvature of Bloch electrons \cite{r22}. Both the extrinsic side jump mechanism and the intrinsic KL mechanism contribution show a quadratic dependence to $\rho_{\text{xx}}$.   
 \newline In ferromagnets, the total Hall resistivity ($\rho_{\text{xy}}$) is a sum of the ordinary Hall resistivity $\rho_{\text{xy}}^{\text{oh}}$ and the anomalous Hall resistivity $\rho_{\text{xy}}^{\text{ah}}$. Fig. 5(a) shows the field driven evolution of the isothermal Hall resistivity curves. It depicts a precipitous growth at low fields that gradually transitions into saturation similar to the field dependent magnetization curves as shown in fig.5(b). This type of behaviour clearly shows the presence of anomalous Hall resistivity in Co$_{8}$Zn$_{8}$Mn$_{4}$. The anomalous Hall resistivity $\rho_{\text{xy}}^{\text{ah}}$ is obtained by extrapolating the linear fit of the high field data to the y axis. Fig 3(b) shows that the $\rho_{\text{xy}}^{\text{ah}}$ tends to decrease with increasing temperature, which is similar to the reported results of Co$_{7}$Zn$_{8}$Mn$_{5}$ \cite{r12} . However, study of Co$_{9}$Zn$_{9}$Mn$_{2}$ has shown conflicting results \cite{r13}  with the anomalous part of resistivity showing a positive change with evolving temperature. 
 The  $\rho_{\text{xy}}^{\text{ah}}$ is expressed as a function of the longitudinal resistivity ($\rho_{\text{xx}}$) as given in the following equation:
\begin{equation}
    \rho_{\text{xy}}^{\text{ah}} =  a\rho_{\text{xx}} +b{\rho^2_{\text{xx}}}
\end{equation}
\begin{figure}[b]
\includegraphics{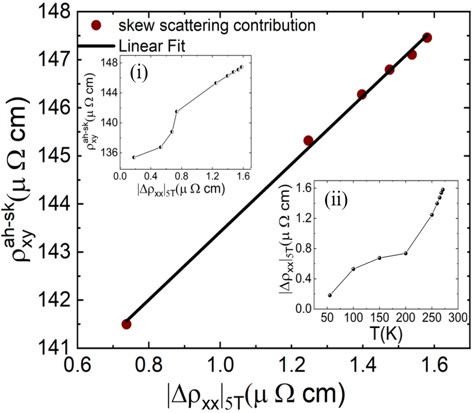}
\caption{\label{fig:epsart} shows the linear fit between the skew scattering contribution of anomalous Hall resistivity ($\rho_{\text{xy}}^{\text{ah-sk}}$) for different temperatures with their corresponding change in resistivity with field ($|\Delta\rho_{\text{xx}}|$$_{5T}$) for T$\ge$200K. Inset (i) shows the plotting between the $\rho_{\text{xy}}^{\text{ah-sk}}$ and $|\Delta\rho_{\text{xx}}|$$_{5T}$ from 50K to 271K. Inset (ii) of (d) shows the variation of $|\Delta\rho_{\text{xx}}|$$_{5T}$ with temperature. }
\end{figure}
                                                                                                                     
Here, a and b are parameters which specifically depend on the material under study. Here, the first term stems from the extrinsic skew scattering mechanism, whereas the second term is an outcome of the extrinsic side jump contribution \cite{r21}  as well as the intrinsic Berry phase mechanism \cite{r22} . Inset of fig.5(b) shows the change in $\rho_{\text{xx}}$ with temperature. Fig.5(c) plots the change in  $\rho_{\text{xy}}^{\text{ah}}$ with  $\rho_{\text{xx}}$. This data has been fitted using the relation  $\rho_{\text{xy}}^{\text{ah}}$= a$\rho_{\text{xx}}$+ b$\rho_{\text{xx}}^2$+c as shown in fig. 4(c). The parameters derived from the fit are a= 1.2216  and b = -0.0057 ($\mu$$\Omega$ cm)$^{-1}$. The negative sign of the quadratic parameter b indicates that the contributions from the side jump mechanism and intrinsic KL mechanism are behaving in the contrary direction as compared to the skew scattering contribution. Here, parameter a is three orders larger as compared to parameter b and this indicates that the  $\rho_{\text{xy}}^{\text{ah}}$ is largely governed by the extrinsic skew scattering contribution. This dominance of skew scattering contribution is in congruence with the results of transport studies done on Co$_{7}$Zn$_{8}$Mn$_{5}$ \cite{r12}. However, the analysis of the AHE in Co$_{9}$Zn$_{9}$Mn$_{2}$ has revealed the dominance of the intrinsic mechanism \cite{r13} . 
\newline Recently, H. Ishizuka et al proposed that the chiral spin fluctuations can give rise to skew scattering contribution in AHE  \cite{r23} . Furthermore, this theory takes into account the Dzyaloshinskii-Moriya (DM) interaction that induces the scalar spin chirality\cite{r23}. Since, this mechanism involves the role of chiral spin fluctuations, it is expected to be more dominant in the high temperature regime \cite{r23}. It is already known that Co$_{8}$Zn$_{8}$Mn$_{4}$ belongs to the chiral space group P4$_{1}$32, where the broken inversion symmetry leads to antisymmetric DM interaction. Hence, we conclude that the skew scattering contribution in Co$_{8}$Zn$_{8}$Mn$_{4}$ stems from the chiral spin fluctuation that leads to this unconventional behaviour.  
\newline Previous theoretical  \cite{r33} and experimental \cite{r34}  reports have suggested that the conventional skew scattering mechanism does not change with temperature. However, the skew scattering contribuion ( $\rho_{\text{ah}}^{sk} $) changes significantly with temperature as shown in the inset of fig.5(c).  As already discussed, the analysis of magnetoresistance behaviour has shown the dominance of spin fluctuations for T$\ge$200K.  These fluctuations get suppressed due to the application of magnetic field resulting in a negative MR.  However, the scattering from spin fluctuations tends to increase with temperature, hence, the MR becomes increasingly negative at high temperatures. Hence, in order to verify the role of spin fluctuations on the skew scattering contribution,  $\rho_{\text{xy}}^{\text{ah-sk}}$ has been plotted with the corresponding change in resistivity under magnetic field $|\Delta\rho_{\text{xx}}|$$_{5T}$ (inset (i) of fig.6). The variation of $|\Delta\rho_{\text{xx}}|$$_{5T}$ with temperature is plotted in inset(ii) of Fig.6. The main panel of fig. 6 shows a linear fit between ,  $\rho_{\text{xy}}^{\text{ah-sk}}$  and $|\Delta\rho_{\text{xx}}|$$_{5T}$ for T $\ge$ 200K. This linear fit clearly depicts that the skew scattering contribution is a result of the spin fluctuations as proposed by Ishizuka et al \cite{r23}. It is noteworthy that there is a deviation from linear fit for T$<$200K where the magnons tend to dominate over the spin fluctuations. This is evidently observed in inset(i) of fig.6.
 \newline Next we discuss the possible reasons for the absence of skew scattering in Co$_{9}$Zn$_{9}$Mn$_{2}$. Although belonging to the same space group as Co$_{8}$Zn$_{8}$Mn$_{4}$ and Co$_{7}$Zn$_{8}$Mn$_{5}$, the only major difference lies in the concentration of Mn ions. Previous reports have shown that the difference in Mn can change the magnetic properties, like the transition temperature \cite{r7}  and the presence or absence of the spin glass phase. It is suggested the Dzyaloshinskii constant D is also affected by the frustration induced by the Mn moments \cite{r35}. All these properties can lead to magnetic complexities in the low Mn family of compounds. Moreover, the Mn atoms occupy the 12d sites which form a 3d hyperkagome network \cite{r25}. Kamil et al have proposed that Kagome and 3D pyrochlore lattices promote chiral spin fluctuations \cite{r36}. So, whether the hyperkagome network promotes chiral spin fluctuations needs to be investigated further that may provide more insight to the absence of skew scattering contribution in Co$_{9}$Zn$_{9}$Mn$_{2}$. 

\section{CONCLUSION}
In conclusion, several aspects of scattering mechanisms are revealed in chiral Co$_{8}$Zn$_{8}$Mn$_{4}$ single crystals through magnetization and magnetotransport measurements. It is found that the scattering from magnons is responsible for the linearity of the MR behaviour at low temperatures. The form of the MR curve deviates from linearity at higher temperatures (T $\ge$ 200 K), indicating the dominance of spin fluctuations over magnons. The conspicuous change in the behaviour of the MR in the paramagnetic phase has been explained by the scattering through local moments using the Khosla and Fischer model. The relationship between the anomalous Hall resistivity and the longitudinal resistivity reveals the dominance of the skew scattering mechanism. The correlation between the skew scattering and the changes in MR confirms the origin of the skew scattering contribution in Co$_{8}$Zn$_{8}$Mn$_{4}$ to be spin fluctuations. Our study determines the dominant scattering mechanisms in low conductivity skyrmionic materials.

\begin{acknowledgments}
PS and PD acknowledge UGC-NET JRF for providing financial support. MS acknowledges CSIR for providing JRF. SP thanks DST-NANOMISSION-CONCEPT for consumables and equipment grants. We are thankful to FIST program of Department of Science and Technology, Government of India for low temperature high magnetic field measurement set-up at JNU. We are grateful to Advanced Instrumentation Research Facility (AIRF), JNU for PPMS measurement facility. 
\end{acknowledgments}

\nocite{*}
\bibliographystyle{IEEEtran}
\bibliography{Co8Zn8Mn4}

\begin{thebibliography}{10}
\providecommand{\url}[1]{#1}
\csname url@samestyle\endcsname
\providecommand{\newblock}{\relax}
\providecommand{\bibinfo}[2]{#2}
\providecommand{\BIBentrySTDinterwordspacing}{\spaceskip=0pt\relax}
\providecommand{\BIBentryALTinterwordstretchfactor}{4}
\providecommand{\BIBentryALTinterwordspacing}{\spaceskip=\fontdimen2\font plus
\BIBentryALTinterwordstretchfactor\fontdimen3\font minus \fontdimen4\font\relax}
\providecommand{\BIBforeignlanguage}[2]{{%
\expandafter\ifx\csname l@#1\endcsname\relax
\typeout{** WARNING: IEEEtran.bst: No hyphenation pattern has been}%
\typeout{** loaded for the language `#1'. Using the pattern for}%
\typeout{** the default language instead.}%
\else
\language=\csname l@#1\endcsname
\fi
#2}}
\providecommand{\BIBdecl}{\relax}
\BIBdecl

\bibitem{r37}
Y.~Tokura and N.~Kanazawa, ``Magnetic skyrmion materials,'' \emph{Chemical Reviews}, vol. 121, no.~5, pp. 2857--2897, 2020.

\bibitem{r1}
S.~M{\"u}hlbauer, B.~Binz, F.~Jonietz, C.~Pfleiderer, A.~Rosch, A.~Neubauer, R.~Georgii, and P.~B{\"o}ni, ``Skyrmion lattice in a chiral magnet,'' \emph{Science}, vol. 323, no. 5916, pp. 915--919, 2009.

\bibitem{r2}
W.~M{\"u}nzer, A.~Neubauer, T.~Adams, S.~M{\"u}hlbauer, C.~Franz, F.~Jonietz, R.~Georgii, P.~B{\"o}ni, B.~Pedersen, M.~Schmidt \emph{et~al.}, ``Skyrmion lattice in the doped semiconductor fe 1- x co x si,'' \emph{Physical Review B}, vol.~81, no.~4, p. 041203, 2010.

\bibitem{r3}
J.~D. Bocarsly, R.~F. Need, R.~Seshadri, and S.~D. Wilson, ``Magnetoentropic signatures of skyrmionic phase behavior in fege,'' \emph{Physical Review B}, vol.~97, no.~10, p. 100404, 2018.

\bibitem{r4}
F.~Qian, L.~J. Bannenberg, H.~Wilhelm, G.~Chaboussant, L.~M. Debeer-Schmitt, M.~P. Schmidt, A.~Aqeel, T.~T. Palstra, E.~Br{\"u}ck, A.~J. Lefering \emph{et~al.}, ``New magnetic phase of the chiral skyrmion material cu2oseo3,'' \emph{Science Advances}, vol.~4, no.~9, p. eaat7323, 2018.

\bibitem{r7}
Y.~Tokunaga, X.~Yu, J.~White, H.~M. R{\o}nnow, D.~Morikawa, Y.~Taguchi, and Y.~Tokura, ``A new class of chiral materials hosting magnetic skyrmions beyond room temperature,'' \emph{Nature communications}, vol.~6, no.~1, p. 7638, 2015.

\bibitem{r8}
J.~D. Bocarsly, C.~Heikes, C.~M. Brown, S.~D. Wilson, and R.~Seshadri, ``Deciphering structural and magnetic disorder in the chiral skyrmion host materials co x zn y mn z (x+ y+ z= 20),'' \emph{Physical Review Materials}, vol.~3, no.~1, p. 014402, 2019.

\bibitem{r9}
K.~Karube, J.~S. White, D.~Morikawa, C.~D. Dewhurst, R.~Cubitt, A.~Kikkawa, X.~Yu, Y.~Tokunaga, T.-h. Arima, H.~M. R{\o}nnow \emph{et~al.}, ``Disordered skyrmion phase stabilized by magnetic frustration in a chiral magnet,'' \emph{Science advances}, vol.~4, no.~9, p. eaar7043, 2018.

\bibitem{r10}
K.~Karube, J.~White, N.~Reynolds, J.~Gavilano, H.~Oike, A.~Kikkawa, F.~Kagawa, Y.~Tokunaga, H.~M. R{\o}nnow, Y.~Tokura \emph{et~al.}, ``Robust metastable skyrmions and their triangular--square lattice structural transition in a high-temperature chiral magnet,'' \emph{Nature materials}, vol.~15, no.~12, pp. 1237--1242, 2016.

\bibitem{r6}
Y.~Tokura and N.~Kanazawa, ``Magnetic skyrmion materials,'' \emph{Chemical Reviews}, vol. 121, no.~5, pp. 2857--2897, 2020.

\bibitem{r38}
M.~Leroux, M.~J. Stolt, S.~Jin, D.~V. Pete, C.~Reichhardt, and B.~Maiorov, ``Skyrmion lattice topological hall effect near room temperature,'' \emph{Scientific reports}, vol.~8, no.~1, p. 15510, 2018.

\bibitem{r11}
N.~Nagaosa, J.~Sinova, S.~Onoda, A.~H. MacDonald, and N.~P. Ong, ``Anomalous hall effect,'' \emph{Reviews of modern physics}, vol.~82, no.~2, p. 1539, 2010.

\bibitem{r12}
H.~Zeng, X.~Zhao, G.~Yu, X.~Luo, S.~Ma, C.~Chen, Z.~Mo, Y.~Zhang, Y.~Chai, J.~Shen \emph{et~al.}, ``Magnetic and transport properties of chiral magnet co7zn8mn5,'' \emph{Journal of Magnetism and Magnetic Materials}, vol. 560, p. 169631, 2022.

\bibitem{r13}
F.~Qi, Y.~Huang, X.~Yao, W.~Lu, and G.~Cao, ``Anomalous electrical transport and magnetic skyrmions in mn-tuned co 9 zn 9 mn 2 single crystals,'' \emph{Physical Review B}, vol. 107, no.~11, p. 115103, 2023.

\bibitem{r14}
M.~Lee, Y.~Onose, Y.~Tokura, and N.~Ong, ``Hidden constant in the anomalous hall effect of high-purity magnet mnsi,'' \emph{Physical Review B}, vol.~75, no.~17, p. 172403, 2007.

\bibitem{r15}
V.~V. Glushkov, I.~I. Lobanova, V.~Y. Ivanov, and S.~V. Demishev, ``Anomalous hall effect in mnsi: Intrinsic to extrinsic crossover,'' \emph{JETP letters}, vol. 101, pp. 459--464, 2015.

\bibitem{r16}
N.~Porter, J.~C. Gartside, and C.~Marrows, ``Scattering mechanisms in textured fege thin films: Magnetoresistance and the anomalous hall effect,'' \emph{Physical Review B}, vol.~90, no.~2, p. 024403, 2014.

\bibitem{r17}
R.~P. Jena, D.~Kumar, and A.~Lakhani, ``Scaling analysis of anomalous hall resistivity in the co2tial heusler alloy,'' \emph{Journal of Physics: Condensed Matter}, vol.~32, no.~36, p. 365703, 2020.

\bibitem{r18}
P.~Saha, M.~Singh, V.~Nagpal, P.~Das, and S.~Patnaik, ``Scaling analysis of anomalous hall resistivity and magnetoresistance in the quasi-two-dimensional ferromagnet fe 3 gete 2,'' \emph{Physical Review B}, vol. 107, no.~3, p. 035115, 2023.

\bibitem{r19}
J.~Smit, ``The spontaneous hall effect in ferromagnetics i,'' \emph{Physica}, vol.~21, no. 6-10, pp. 877--887, 1955.

\bibitem{r20}
------, ``The spontaneous hall effect in ferromagnetics ii,'' \emph{Physica}, vol.~24, no. 1-5, pp. 39--51, 1958.

\bibitem{r21}
L.~Berger, ``Side-jump mechanism for the hall effect of ferromagnets,'' \emph{Physical Review B}, vol.~2, no.~11, p. 4559, 1970.

\bibitem{r22}
R.~Karplus and J.~Luttinger, ``Hall effect in ferromagnetics,'' \emph{Physical Review}, vol.~95, no.~5, p. 1154, 1954.

\bibitem{r23}
H.~Ishizuka and N.~Nagaosa, ``Spin chirality induced skew scattering and anomalous hall effect in chiral magnets,'' \emph{Science advances}, vol.~4, no.~2, p. eaap9962, 2018.

\bibitem{r24}
D.~Zhang, H.~Ishizuka, N.~Lu, Y.~Wang, N.~Nagaosa, P.~Yu, and Q.-K. Xue, ``Anomalous hall effect and spin fluctuations in ionic liquid gated srcoo 3 thin films,'' \emph{Physical Review B}, vol.~97, no.~18, p. 184433, 2018.

\bibitem{r25}
J.~D. Bocarsly, C.~Heikes, C.~M. Brown, S.~D. Wilson, and R.~Seshadri, ``Deciphering structural and magnetic disorder in the chiral skyrmion host materials co x zn y mn z (x+ y+ z= 20),'' \emph{Physical Review Materials}, vol.~3, no.~1, p. 014402, 2019.

\bibitem{r28}
T.~Nakajima, K.~Karube, Y.~Ishikawa, M.~Yonemura, N.~Reynolds, J.~White, H.~R{\o}nnow, A.~Kikkawa, Y.~Tokunaga, Y.~Taguchi \emph{et~al.}, ``Correlation between site occupancies and spin-glass transition in skyrmion host co 10- x 2 zn 10- x 2 mn x,'' \emph{Physical Review B}, vol. 100, no.~6, p. 064407, 2019.

\bibitem{r26}
M.~E. Henderson, J.~Beare, S.~Sharma, M.~Bleuel, P.~Clancy, D.~G. Cory, M.~G. Huber, C.~A. Marjerrison, M.~Pula, D.~Sarenac \emph{et~al.}, ``Characterization of a disordered above room temperature skyrmion material co8zn8mn4,'' \emph{Materials}, vol.~14, no.~16, p. 4689, 2021.

\bibitem{r27}
Y.~Bitla, S.~Saha, A.~K. Patra, G.~Basheed \emph{et~al.}, ``Chiral-fluctuations mediated helical to paramagnetic phase transition and scaling study in $\beta$-mn type co8zn8mn4 chiral magnet,'' \emph{Journal of Physics: Condensed Matter}, vol.~35, no.~17, p. 175801, 2023.

\bibitem{r29}
M.~M.~H. Polash and D.~Vashaee, ``Spin fluctuations yield zt enhancement in ferromagnets,'' \emph{Iscience}, vol.~24, no.~11, 2021.

\bibitem{r30}
B.~Raquet, M.~Viret, E.~Sondergard, O.~Cespedes, and R.~Mamy, ``Electron-magnon scattering and magnetic resistivity in 3 d ferromagnets,'' \emph{Physical Review B}, vol.~66, no.~2, p. 024433, 2002.

\bibitem{r31}
S.~Kaul, ``Spin-wave and spin-fluctuation contributions to the magnetoresistance of weak itinerant-electron ferromagnets,'' \emph{Journal of Physics: Condensed Matter}, vol.~17, no.~36, p. 5595, 2005.

\bibitem{r32}
R.~Khosla and J.~Fischer, ``Magnetoresistance in degenerate cds: Localized magnetic moments,'' \emph{Physical Review B}, vol.~2, no.~10, p. 4084, 1970.

\bibitem{r33}
A.~Cr{\'e}pieux and P.~Bruno, ``Theory of the anomalous hall effect from the kubo formula and the dirac equation,'' \emph{Physical Review B}, vol.~64, no.~1, p. 014416, 2001.

\bibitem{r34}
Y.~Tian, L.~Ye, and X.~Jin, ``Proper scaling of the anomalous hall effect,'' \emph{Physical review letters}, vol. 103, no.~8, p. 087206, 2009.

\bibitem{r35}
V.~Ukleev, K.~Pschenichnyi, O.~Utesov, K.~Karube, S.~M{\"u}hlbauer, R.~Cubitt, Y.~Tokura, Y.~Taguchi, J.~White, and S.~Grigoriev, ``Spin wave stiffness and damping in a frustrated chiral helimagnet co 8 zn 8 mn 4 as measured by small-angle neutron scattering,'' \emph{Physical Review Research}, vol.~4, no.~2, p. 023239, 2022.

\bibitem{r36}
K.~K. Kolincio, M.~Hirschberger, J.~Masell, T.-h. Arima, N.~Nagaosa, and Y.~Tokura, ``Kagom{\'e} lattice promotes chiral spin fluctuations,'' \emph{Physical Review Letters}, vol. 130, no.~13, p. 136701, 2023.

\bibitem{r5}
Z.~Ma and L.~Li, ``Recent progress in exploring the magnetic solids with room-temperature skyrmionic spin configurations,'' \emph{physica status solidi (a)}, vol. 219, no.~1, p. 2100333, 2022.

\end{thebibliography}

\end{document}